\documentclass[preprint,12pt]{elsarticle}
\usepackage{graphicx}
\usepackage{amssymb}
\usepackage{amsmath}
\usepackage[utf8x]{inputenc} 
\usepackage{epstopdf}
\usepackage{natbib}
\usepackage{bm}
\usepackage{siunitx}
\usepackage[misc]{ifsym}
\usepackage{subfigure}
\usepackage{xcolor}
\usepackage[colorlinks=true,linkcolor=blue,anchorcolor=blue,citecolor=blue]{hyperref} 
\usepackage{booktabs}
\usepackage{multirow}
\usepackage{lipsum}

\usepackage{lineno,hyperref}
\usepackage[symbol]{footmisc}
\modulolinenumbers[1]

\usepackage{color}
\usepackage{ulem}

\journal{Int. J. Heat Mass Transfer}

\begin{document}

\begin{frontmatter}

\title{Thermocapillary flow transition in an evaporating liquid layer in a heated cylindrical cell} 

\author{Wenjun Liu$^{a,b,d}$, Paul G. Chen$^b$\footnote{Corresponding author. Electronic addresse: gang.chen@univ-amu.fr}, Jalil Ouazzani$^{c,e}$, Qiusheng Liu$^{a,d}\footnote{Corresponding author. Electronic addresse: liu@imech.ac.cn}$} 

\address{ 
$^a$ Institute of Mechanics, Chinese Academy of Sciences, Beijing, China\\
$^b$ Aix Marseille University, CNRS, Centrale Marseille, M2P2, Marseille, France\\
$^c$ Arcofluid Consulting LLC, 309 N Orange Ave, Orlando, FL 32801, USA\\
$^d$ School of Engineering Science, University of Chinese Academy of Sciences, Beijing, China\\
$^e$ Guilin University of Electronic Technology, Guilin, China}

\begin{abstract}

Motived by recent ground-based and microgravity experiments investigating the interfacial dynamics of a volatile liquid (FC-72, $Pr=12.34$) contained in a heated cylindrical cell, we numerically study the thermocapillary-driven flow in such an evaporating liquid layer. Particular attention is given to the prediction of the transition of the axisymmetric flow to fully three-dimensional patterns when the applied temperature increases. The numerical simulations rely on an improved one-sided model of evaporation by including heat and mass transfer through the gas phase via the heat transfer Biot number and the evaporative Biot number. We present the axisymmetric flow characteristics, show the variation of the transition points with these Biot numbers, and more importantly elucidate the twofold role of the latent heat of evaporation in the stability; evaporation not only destabilizes the flow but also stabilizes it, depending upon the place where the evaporation-induced thermal gradients come into play. We also show that buoyancy in the liquid layer has a stabilizing effect, though its effect is insignificant. At high Marangoni numbers, the numerical simulations revealed smaller-scale thermal patterns formed on the surface of a thinner evaporating layer, in qualitative agreement with experimental observations. The present work helps to gain a better understanding of the role of a phase change in the thermocapillary instability of an evaporating liquid layer.

\begin{keyword}
Thermocapillary convection; Flow transition; Evaporation; Numerical simulation.
\end{keyword}

\end{abstract}

\end{frontmatter}


\section{Introduction}\label{sec:intro}
A thin fluid layer open to the air is unstable when heated from below or cooled from above due to evaporation if the Rayleigh number ($Ra$) exceeds the primary threshold $Ra_c$ for purely buoyancy-driven convection, or if the Marangoni number ($Ma$) exceeds the primary threshold $Ma_c$ for purely surface-tension-driven convection. These situations with the combined effects of buoyancy and surface tension -- referred to as the Rayleigh-B\'enard-Marangoni convection instability, arise in a wide range of problems of fundamental and practical importance (see e.g., recent reviews by Fauve~\cite{Fauve_2017}; Gallaire and Brun~\cite{Gallaire_2017}).

When the liquid-gas systems are confined by a heated or cooled sidewall relative to the surrounding environment, lateral temperature gradients can result and induce surface-tension-driven flow near the sidewall. It is, therefore, useful to distingush two classes of instabilities, depending upon whether the applied temperature gradient is primarily normal to the interface (Marangoni-convection instability) or along the interface (thermocapillary-convection instability), see the review article by Schatz and Neitzel \cite{schatz2001experiments}.

For Marangoni-convection instability, linear-stability analysis of the conductive or basic state (i.e., motionless) predicts two different types of primary instabilities that may arise in experiments, that is, a short-wavelength instability for thick liquid layers and a long-wavelength instability for sufficiently thin liquid layers. Cellular patterns observed in experiments at convective onset are due to the short-wavelength models~\cite{schatz1995onset,cerisier1996topological}. In contrast to the Marangoni convection where there exists a critical applied temperature gradient (normal to the interface), thermocapillary convection appears whenever a surface-temperature gradient exists, no matter how small. The problem of stability of these basic (steady convective) states is rarely analytically tractable in practice because the basic states themselves may be unavailable  or too complex to deal with analytically. Prior research on thermocapillary instabilities has been mostly focused on three configurations due to their relevance to crystal-growth application, namely, liquid bridge {\cite{preisser1983liquidbridge,Shevtsova_1998}, rectangular layer~\cite{smith1983instabilities,Riley_1998} or slot~\cite{gillon1996combined}, and annular geometry~\cite{kamotani1992experimental}. 
 
Buoyancy-driven motion and thermocapillary convection can operate simultaneously in terrestrial experiments, and the relative strength of the two mechanisms is given by the ratio $Ra/Ma$ called the dynamic Bond number ($Bd$). Since $Bd \sim gh^2$, with $g$ being the gravitational acceleration and $h$ the thickness of liquid layer, the condition for thermocapillarity to dominate over buoyancy in experiments is easily satisfied by choosing $h$ sufficiently small or by carrying out experiments under microgravity conditions where $g$ almost vanishes~\cite{mg2006,Hu_2014,mg2018,watanabe2018terrestrial}.

Most previous experiments considered that the liquid-gas interface is adiabatic, but the heat and mass transfer through the interface has a significantly impact on the flow stability.  As such, Li et al. \cite{li2012effect} introduced Newton's cooling law via the Biot number to represent the heat transfer at the interface. Liu and Kabov \cite{liu2012instabilities} added the evaporation effect on the convective instabilities in a horizontal liquid layer. Li et al. \cite{li2014experimental} and Roman et al. \cite{grigoriev2018effect} used experimental and numerical methods to investigate the effects of gas-phase transport on the critical conditions of thermal convection in an evaporating liquid layer. Wei and Duan \cite{duanfei2018} investigated the onset of long-wave-instability in an evaporating  liquid layer subjected to vapor recoil, thermocapillarity, gravity, and ambient cooling.

{\color{black}
A thin liquid layer with evaporation plays a significant role in various physical fundamental problems and industrial utilization including thin-film coating \cite{cummings2018modeling,kim2004highly}, organic electronics \cite{agar2016highly,shao2014understanding} and heat-energy engineering \cite{peng2018low,chen2017magnetically}. In terms of the industrial application technology of crystal preparation, it is crucial to study the flow stability in shallow liquid layers, see the review article by Imaishi and Kakimoto \cite{imaishi2002convective}. Additionally, unlike large-scale ground fluid systems, the surface tension-driven interfacial flow becomes the foremost factor affecting the fluid heat and mass transfer process in a microgravity environment \cite{mg2006,Hu_2014}. The research of the thermocapillary convection stability and heat mass transfer law in an evaporating thin liquid layer will provide the necessary theoretical basis for the design and development of thermal fluid facilities, such as space heat pipes, fluid-on-orbit management, and space life support systems. The present work is a part of numerical studies of the space project of the two-phase fluid experiment, which is one of the microgravity fluid physics experiments scheduled onboard Chinese Tianzhou-1 cargo ship \cite{yidong2014science,yidong2016space}. The reason we have selected FC-72 as a working liquid is as follows. Due to a large number of electronic components in our space experiment equipment, from the perspective of space experiment safety, the experimental working medium should be dielectric and non-flammable. FC-72 liquid is non-conductive, non-flammable, non-toxic, and resistant to pollution, which meets the safety design principles of space experiments. Furthermore, from the perspective of flow stability, the surface tension of FC-72 is relatively low that a small temperature difference can drive the Marangoni convection in the liquid, which is helpful to obtain physical phenomena in both ground and space experiments. Also, because of its high heat exchange efficiency, FC-72 is often used as an experimental working fluid in heat pipes for space applications.
}

{\color{black} 
The present study is also motivated by recent ground-based experiments~\cite{TZ} investigating the dynamics of an evaporating layer of FC-72 liquid exposed to the air. The liquid layer is contained in a shallow cylindrical pool, heated both from below and from the side at a fixed higher temperature relative to its environment. An overhead infrared camera is employed to observe the surface temperature field and the development of the flow pattern in the evaporating layer.}
The experiments help to gain a better understanding of the physics of the thermocapillary instability associated with evaporative phase change. Even in the absence of gravity, the thermocapillary effect is present, so there doesn't exist the onset of convection. One would expect a steady axisymmetric flow induced by thermocapillarity and buoyancy (in the presence of gravity) if the imposed temperature difference is sufficiently small. Before going any further, the primary question of interest concerns the stability of the steady basic state. 

To the best our knowledge, an evaporating liquid layer subjected to a bidirectional temperature gradient has rarely been studied in detail experimentally and/or numerically. The present work aims to predict the stability threshold of the axisymmetric flow undergoing a transition to a fully 3D flow pattern, and to study how heat and mass transfer across the surface (i.e., liquid-gas interface) affects the transition point, namely the critical Marangoni number $Ma_c$. To this end, we perform fully three-dimensional computer simulations based on a one-sided evaporation model in which vapor dynamics is ignored and the heat and mass fluxes across the interface are represented by global transfer coefficients. 

The paper is structured as follows. In Section~\ref{Sec:Math}, we present the mathematical model and basic underlying assumptions. A brief description of the numerical method used and a mesh convergence study are given in Section~\ref{Sec:Simu}. Simulation results are presented and discussed in Section~\ref{Sec:Results}, with an emphasis on the transition point -- the critical Marangoni number $Ma_c$, and the effects of various control parameters on $Ma_c$.  A summary of the main results of the present work and concluding remarks are given in Section~\ref{Sec:Conclusion}.

\section{Problem formulation}\label{Sec:Math}
\subsection{Basic assumptions}

We consider a viscous liquid layer enclosed in a cylindrical cell, as depicted in Fig. \ref{fig1}. The liquid layer is an incompressible Newtonian fluid with constant material properties and surrounded by a passive gas, whose viscosity and thermal conductivity are taken to be very small compared to those of the liquid. The layer is subjected to a bidirectional temperature gradient due to a uniformly heated ($T_h$) bottom wall and sidewall, compared to the far-field gas temperature ($T_\infty$). Even in the absence of gravity, thermocapillarity induces fluid flow in the layer. This flow is coupled to the buoyancy-driven motion in the presence of gravity.  Besides, the layer is evaporating, at a rate being dependent on the volatility of working fluid. In the present work, the working liquid is Fluorinert liquid FC-72, which is a clear, colorless, fully-fluorinated liquid. Its thermophysical properties are listed in Table \ref{tab1}, in which $\rho$, $\mu$, $\nu$, $\alpha$, $\kappa$, $\beta$, $\sigma$, and $\mathcal{L}$ respectively denote fluid density, dynamic viscosity, kinematic viscosity, thermal diffusivity, thermal conductivity, thermal expansion coefficient, surface tension, and the latent heat of vaporization.  The surface tension is assumed to be a decreasing function of temperature $T$, i.e., $\sigma(T) = \sigma_0(T_\infty) - \gamma (T - T_\infty)$, where $\gamma$ ($\equiv - \mbox{d} \sigma/\mbox{d} T >0$) is the (negative) temperature coefficient of surface tension $\sigma$. These symbols are to be used straightforwardly hereafter without specifying again their meanings. 
  
\begin{figure}[htbp]
	\centering
	\includegraphics[width=0.5\textwidth]{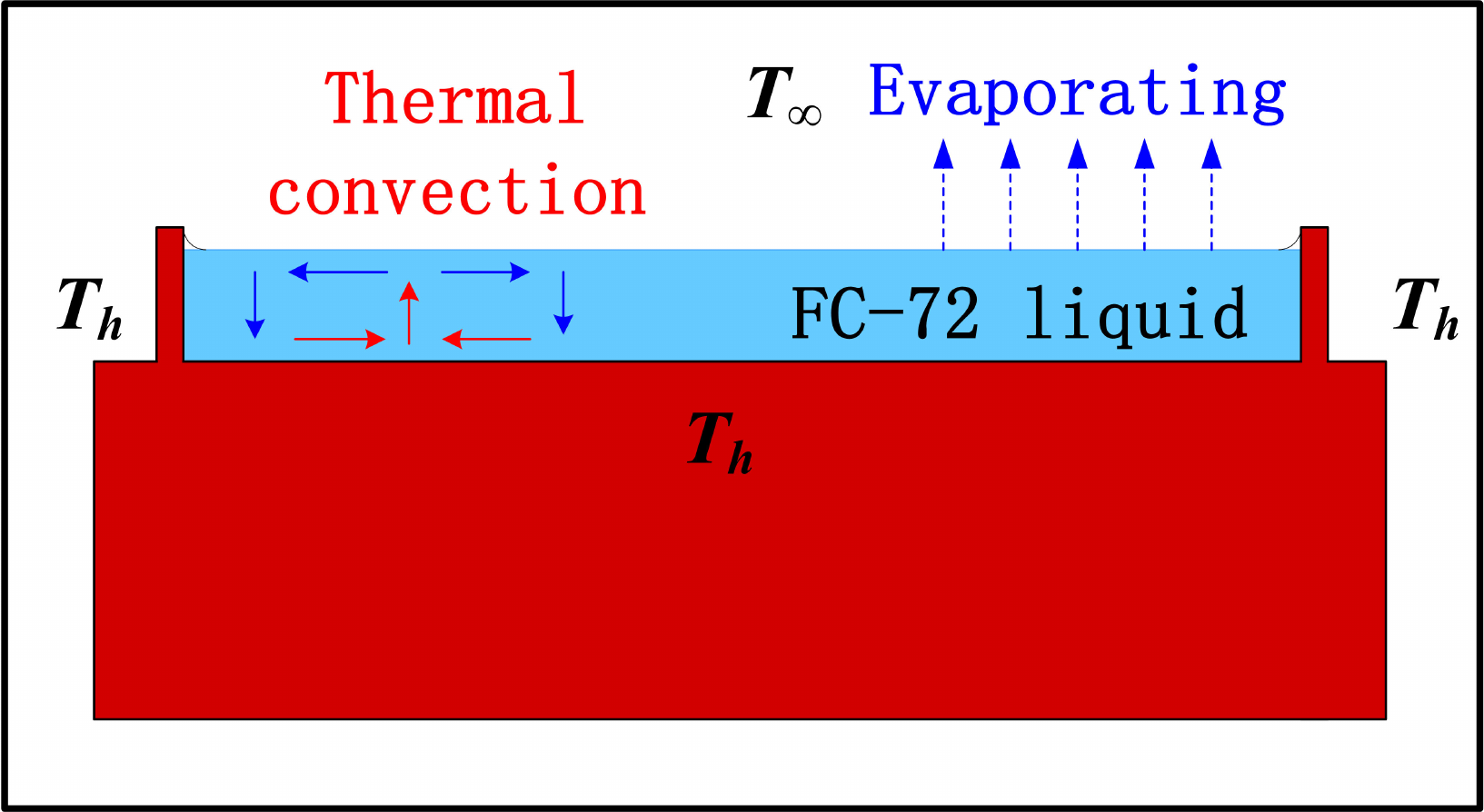}
	\caption{Schematic depiction of the experimental setup. The liquid layer is heated from below and from the side. \label{fig1}}
\end{figure}

\begin{table*}[htbp]
\scriptsize
\centering
\caption{Thermophysical properties of FC-72 at $298.15$ K and 1 atm used for calculations. A value of $h=2$ mm is used in calculating the dimensionless groups. \label{tab1}}
\setlength{\leftskip}{-30pt}
\begin{tabular}{ccccccccc}
\toprule
$\rho$ $\times$ 10$^{-3}$ & $\mu$ $\times$ 10$^{4}$ & $\nu$ $\times$ 10$^{7}$ & $\alpha$ $\times$ 10$^{8}$ & $\kappa$ $\times$ 10$^{2}$ & $\beta$ $\times$ 10$^{3}$ & $\sigma_0$ $\times$ 10$^{2}$ & $\gamma$ $\times$ 10$^{5}$ & $\mathcal{L}$ $\times$ 10$^{-4}$\\
(kg~m$^{-3}$) & (kg~m$^{-1}~$s$^{-1}$) & (m$^2$~s$^{-1}$) & (m$^2$ s$^{-1}$) & (W m$^{-1}$~K$^{-1}$) & (K$^{-1}$) & (N m$^{-1}$) & (N m$^{-1}~$K$^{-1}$) & (J~kg$^{-1}$) \\
\midrule
1.68 & 6.38 & 3.80 & 3.08 & 5.70 & 1.56 & 1.06 & 8.58 & 8.80 \\
\bottomrule
\end{tabular}%
\end{table*}%

As shown in Fig.~\ref{fig2}, the mathematical model of the motion of a layer evaporating into a passive medium is based on the one-sided model of evaporation (negligible vapor density and viscosity). Three simplifying assumptions are additionally made:  (i) the liquid layer remains planar, (ii) its thickness $h$ remains constant, and (iii)  the heat and mass fluxes cross the interface are described by global heat and mass transfer coefficients. We note that several previously reported studies on Rayleigh-B\'enard-Marangoni convection under evaporation used such a one-sided model (e.g.,~\cite{ozen2004physics,doumenc2010transient,liyourong2019}).

\begin{figure}[htbp]
	\centering
	\includegraphics[width=0.5\textwidth]{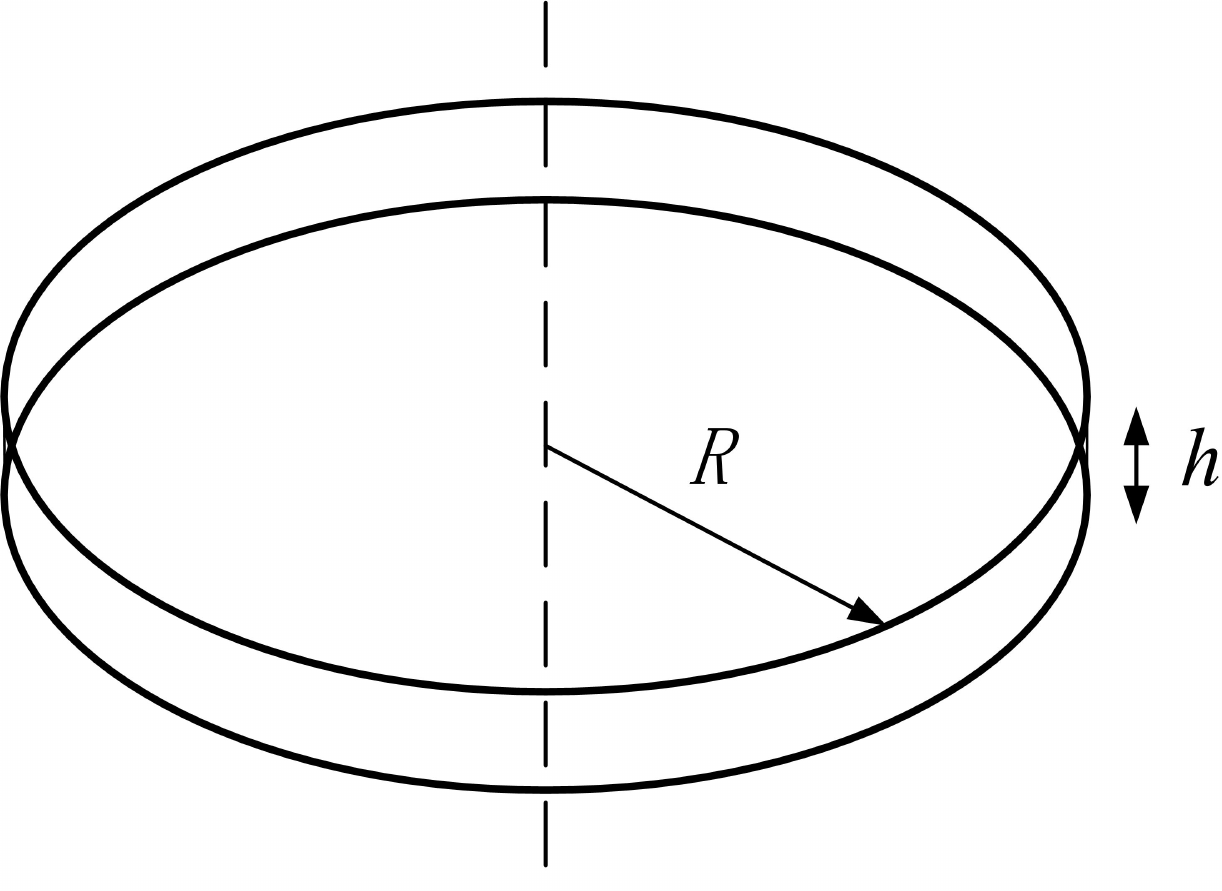}
	\caption{A simplified one-sided model of evaporation. A thin liquid layer is enclosed in a cylindrical cell of radius $R$ ($R=10$ mm, $h=2$ mm). The bottom wall and the sidewall in contact with the liquid are maintained at  temperature $T_h$ ($\Delta T = T_h - T_\infty > 0$).   \label{fig2}}
\end{figure}

Doumenc et al.~\cite{doumenc2010transient} discussed the conditions under which the first two assumptions can be adopted. Firstly, surface deformation can be neglected on scales of order of $h$ if the crispation number $Cr$ ($\equiv \mu\alpha/(\sigma_0 h)$) $\ll 1$ and the Galileo number $Ga$ ($\equiv g h^3/(\nu \alpha)$) $\gg 1$. These conditions are indeed fulfilled herein since $Cr=9.3\times 10^{-7}$ and $Ga=6.7\times 10^6$ (see Table~\ref{tab1}). Secondly, if the P\'eclet number $Pe$ ($\equiv U_\mathrm{ev}/U$) $\ll 1$, the surface displacement $h'$ remains negligible compared to the thickness of the layer $h$. Here, we estimate the interface velocity due to evaporation, $U_\mathrm{ev} = \kappa \Delta T/(\rho \mathcal{L} h)$,  from a balance of heat transfer and latent heat terms in the interfacial energy balance~\cite{Chen_2017a}. A characteristic velocity scale, $U = \gamma \Delta T/\mu$,  is obtained by a balance between viscous stress and thermocapillarity along the liquid-gas interface~\cite{Chen_2017b}. From Table~\ref{tab1}, we get an estimate of $Pe$ ($=h'/h$) $\approx 1.4 \times 10^{-6}$, thereby justifying the second assumption.

Finally, instead of using the energy balance and mass conservation conditions at the free surface, as in, e.g.,~\cite{Chen_2017b}, we adopt a phenomenological approach in which the heat flux and mass conservation across the interface are given by global heat and mass transfer coefficients. As such, the heat transfer in the gas phase and the cooling effect due to evaporation can be described respectively by an equivalent Biot number~\cite{Burelback_1988,ozen2004physics,doumenc2010transient,Karapetsas_2012}. Since our main concern lies more in the prediction of the instability thresholds of the basic state and than the detailed account of the transfer, the last assumption allows us to greatly simplify the description of the coupling conditions at the free surface while retaining the physics of the problem under study. 

\subsection{Governing equations}
The flow and heat transfer in the liquid layer is governed by the incompressible Navier-Stokes equations, 
continuity and energy equation. We formulate the governing equations in a cylindrical coordinate system ($\bm{e}_r, \bm{e}_\phi, \bm{e}_z$), where the bottom of the layer is located at $z=0$ and the upper free surface at $z=h$. Under the Boussinesq approximation,  the velocity field $\bm{u}$~($\equiv u_r\bm{e}_r+u_\phi \bm{e}_\phi+u_z\bm{e}_z$), pressure field $p$, and temperature field $T$ satisfy 

\begin{subequations} \label{Eq:A}
\begin{gather}
\frac{\partial \bm{u}}{\partial t}+ \left (\bm{u} \bm{\cdot} \bm{\nabla}\right ) \bm{u}=-\frac{1}{\rho} \bm{\nabla} p+\nu \nabla^2 \bm{u}+g\beta(T-T_{\infty})\bm{e}_z,  \\
\bm{\nabla} \bm{\cdot} \bm{u} = 0 , \label{Eq:A1} \\
\frac{\partial T}{\partial t}+\left (\bm{u} \bm{\cdot} \bm{\nabla} \right ) T =\alpha \nabla^2T.
\end{gather}
\end{subequations} 
Here, the density $\rho$ is taken to be the density at $T=T_{\infty}$. These equations are subject to appropriate boundary conditions that are specified hereafter.

At the bottom wall and the sidewall, the velocity satisfies the no-slip condition and the temperature
is maintained at temperature $T_h$,
\begin{equation}
\bm{u}= \bm{0}, \quad T=T_h \quad \mbox{at} \quad z=0 \quad \mbox{and} \quad  r=R.
\end{equation} 

At the upper free surface, 
the abovementioned first two assumptions (i.e., planar and constant layer)
lead to a simplified kinematic boundary condition,
\begin{equation}
u_z= 0 \quad \mbox{at} \quad  z=h,
\end{equation}
and a reduced form of the dynamic boundary condition for the tangential stress balance,
\begin{equation}
\mu \frac{\partial u_r}{\partial z}= - \gamma \frac{\partial T}{\partial r},\quad \mu \frac{\partial u_\phi}{\partial z}= - \gamma \frac{\partial T}{r\partial \phi} \quad \mbox{at} \quad z=h. \label{eq:dym}
\end{equation}
Here, the shear viscous stress (left-hand side term) is balanced with the Marangoni stress (right-hand side term) due to the temperature dependence of the surface tension. 

The problem formulation is completed by specifying appropriate boundary conditions for the temperature at the free surface, which constitutes a key ingredient of the one-sided model of evaporation. Fundamentally, the energy balance implies that the heat flux across the interface experiences discontinuity due to the latent heat of vaporization $\mathcal{L}$,
\begin{equation}
q_l - q_v =j_m \mathcal{L} \quad \mbox{at} \quad  z=h.
\end{equation}
Here, $q_l$ ($\equiv - \kappa \partial T/\partial z$) represents the heat flux in the liquid phase, and $q_v$ ($\equiv - \kappa_v \partial T/\partial z$) represents the heat flux in the gas phase, which is modeled using Newton's law involving the heat transfer coefficient $h_\mathrm{th}$ (in W m$^{-2}$ K$^{-1}$). The heat flux is proportional to the difference in temperature between the surface and ambient medium far from the surface, 
\begin{equation}
q_v = h_\mathrm{th} (T_s - T_\infty). 
\end{equation}

The evaporative mass flux of vapor from the layer $j_m$ (in kg~m$^{-2}$~s$^{-1}$) is modeled in a similar manner. It depends on the saturation pressure, the vapor diffusivity, and the far-field concentration:
\begin{equation}
j_m  \equiv  - D \partial c/\partial z  = h_m (c_\mathrm{sat}(T_s) - c_\infty), 
\end{equation}
where $D$ denotes the mass diffusivity of vapor in the ambient air, $c$ is the mass concentration of the liquid vapor, $c_\mathrm{sat}$ and $c_\infty$ represent the vapor concentration in the gas-phase near the surface and far from the surface, respectively, and $h_m$ is the phenomenological mass transfer coefficient in the gas (in m s$^{-1}$). The mass flux $j_m$ is readily converted into a heat flux. To that, we use the evaporative resistance model~\cite{GIUSTINI_2016}, we then have
\begin{equation} \label{eq:jm}
j_m \mathcal{L} = h_\mathrm{ev} (T_s - T_\infty), 
\end{equation}
where $h_\mathrm{ev}$ is the evaporative heat transfer coefficient. Giustini et al.~\cite{GIUSTINI_2016} provided an estimate for its value. It depends on the fluid properties, and in particular on the evaporation coefficient and the latent heat of evaporation. We shall estimate $h_\mathrm{ev}$ based on our ground evaporation experiments~\cite{TZ}.

Putting them together, we obtain the final expression for this one-sided model:
\begin{equation}
\kappa \frac{\partial T}{\partial z} + h_\mathrm{th} (T - T_\infty) + h_\mathrm{ev}(T - T_\infty)=0 \quad \mbox{at} \quad  z=h.
\end{equation}
While it is possible to use a global transfer coefficient by combining the two heat transfer coefficients, we distinguish the two terms here to examine their contributions separately.

We now nondimensionalize the governing equations and boundary conditions. Here, it is natural to use the thickness of the liquid layer $h$ as the length scale, a characteristic surface-tension-driven velocity $U$ ($=  \gamma \Delta T/\mu$) for the velocity scale,  $h/U$ for the time scale. The nondimensionalization leads to equations for the velocity field $\bm{u}$ and temperature $\theta$ ($\equiv (T - T_\infty)/\Delta T$) (all variables and parameters are henceforth dimensionless):\par
\begin{subequations} \label{Eq:AB}
\begin{gather}
\frac{\partial \bm{u}}{\partial t}+ \left (\bm{u} \bm{\cdot} \bm{\nabla}\right ) \bm{u}=- \bm{\nabla} p+\frac{Pr}{Ma} \nabla^2 \bm{u}+Bd\frac{Pr}{Ma}\theta\,\bm{e}_z, \\
\bm{\nabla} \bm{\cdot} \bm{u} = 0, \\
\frac{\partial \theta}{\partial t}+\left (\bm{u} \bm{\cdot} \bm{\nabla} \right ) \theta =\frac{1}{Ma} \nabla^2 \theta, \\
\frac{\partial u_r}{\partial z}+\frac{\partial \theta}{\partial r}=  \frac{\partial u_\phi}{\partial z}+ \frac{\partial \theta}{r\partial \phi}=0 \quad \mbox{at} \quad z=1,\\
\frac{\partial \theta}{\partial z}+Bi \,\theta+B_\mathrm{ev}\theta=0 \quad \mbox{at} \quad z=1, \label{Eq:AB5}\\
u_z=0\quad \mbox{at} \quad z=1,\\
u_r=u_\phi=u_z=0, \quad \theta=1 \quad \mbox{at} \quad z=0, \\
u_r=u_\phi=u_z=0, \quad \theta=1 \quad \mbox{at} \quad r=5.
\end{gather}
\end{subequations} 
Hence, the flow in the liquid layer is determined by five independent dimensionless parameters: the Marangoni number $Ma$, the dynamic Bond number $Bd$ ($\equiv Ra/Ma$, instead of using the Rayleigh number $Ra$), the Prandtl number $Pr$, the (heat transfer) Biot number $Bi$, and the evaporative Biot number $B_\mathrm{ev}$: \par
\begin{equation*}
\begin{split}
Ma&=\frac{\gamma \Delta T h}{\mu \alpha} , \quad Bd=\frac{\rho g \beta  h^2}{\gamma} , \quad Pr=\frac{\nu}{\alpha},
\end{split}
\end{equation*}
\begin{equation*}
\begin{split}
Bi=\frac{h_\mathrm{th}h}{\kappa}, \quad B_\mathrm{ev} = \frac{h_\mathrm{ev} h}{\kappa}.
\end{split}
\end{equation*}

We shall note that for our particular working fluid FC-72 (see Table~\ref{tab1}), we have $Pr=12.34$. We study two scenarios for a layer of 2 mm thick to simulate microgravity and normal gravity conditions, respectively: one is 0g, i.e., $Bd=0$, and the other is 1g, i.e., $Bd=1.2$. We also note that in Eq.~(\ref{Eq:AB5}) the term $Bi\,\theta$ represents the heat transfer in the gas phase and the term $B_\mathrm{ev}\theta$ describes the cooling effect due to evaporation. While we may define a global Biot number by combining the two Biot numbers, we retain the two Biot numbers to examine their individual effects.  Another reason is that in a well-controlled physical experiment heat loss to the ambient gas is generally small, meaning a small Biot number, usually $Bi < 1$. A value of $Bi=0$ represents a perfectly insulated interface. By contrast, the evaporative Biot number is relatively large. 

{\color{black}
\begin{figure}[htbp]
	\centering
	\includegraphics[width=0.7\textwidth]{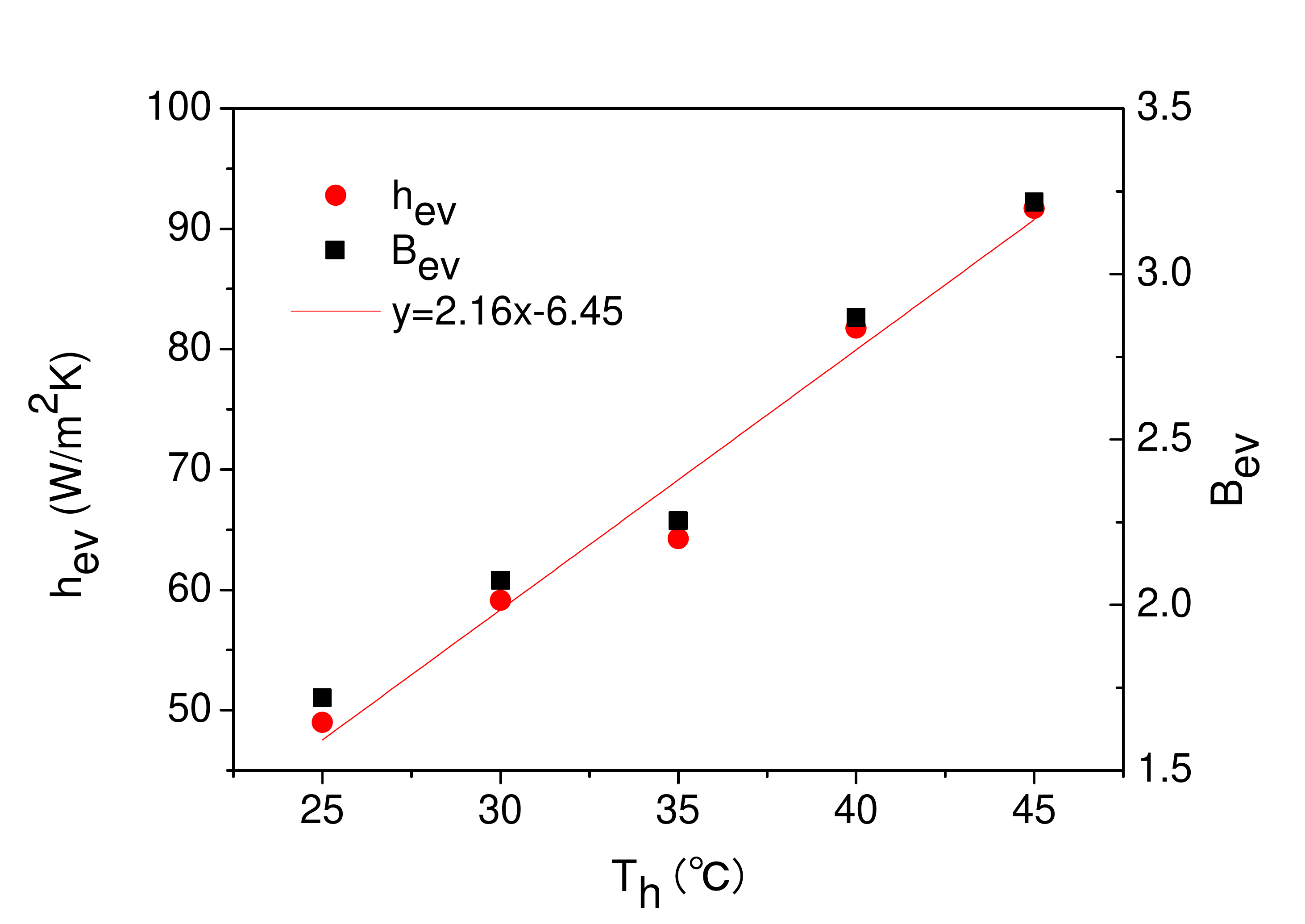}
	\caption{{\color{black}Estimated evaporative heat transfer coefficient $h_\mathrm{ev}$ and evaporative Biot number $B_\mathrm{ev}$ as a function of the applied substrate temperature $T_h$ (at $T_\infty=$ \SI{25}{\degreeCelsius}). Experimental data of the average evaporation rate are taken from Ref.~\cite{TZ}. }}\label{figa3}
\end{figure}
We use the average evaporation rate obtained in Ref.~\cite{TZ} (their figure 4) to estimate the evaporative heat transfer coefficient $h_\mathrm{ev}$ and the evaporative Biot number $B_\mathrm{ev}$. Specifically, equation~(\ref{eq:jm}) allows for an estimate of $h_\mathrm{ev}$ for a given average evaporation rate if we assume $T_s -T_\infty \approx T_h - T_\infty$ (which is not true since $T_s < T_h$). In this sense, the value of $h_\mathrm{ev}$ is underestimated  to some extent. Given that we are at the lower temperature range (i.e., 25$\sim$\SI{35}{\degreeCelsius}) and according to Fig.~\ref{figa3}, it is a reasonable assumption if we set $B_\mathrm{ev}\approx 2.0$.}
Hence, in simulations, we vary the Biot number from 0.05 to 1 while setting $B_\mathrm{ev}=0$ without evaporation or $B_\mathrm{ev}=2$ with evaporation. As such, for a given set of dimensionless groups (i.e., $Bd$, $Bi$, and $B_\mathrm{ev}$) the only control parameter becomes the Marangoni number $Ma$, which corresponds to an actual control parameter in physics experiments -- the applied temperature difference $\Delta T$.  Our main goal is to elucidate the features of the flow and examine how the flow pattern changes with $Ma$, and particularly determine the critical Marangoni number $Ma_c$ beyond which the basic flow undergoes a transition to  3D pattern. 

\section{Numerical simulation}\label{Sec:Simu}

\subsection{Simulation code}

Numerical simulations are performed in the cylindrical coordinate system $(r, \phi, z)$ using PHOENICS 2018~\cite{Phoenics_2018,Clus_2009}, a commercial computational fluid dynamics (CFD) software. PHOENICS is a general-purpose CFD code simulating steady or unsteady, two-or three-dimensional turbulent or laminar, single-phase or multi-phase, compressible or incompressible flows. 
The numerical procedure is of the finite-volume type in which the original partial differential equations are converted into algebraic finite-volume equations with the aid of discretization assumptions for the transient, convection, diffusion and source terms. For that purpose, the solution domain is subdivided into several control volumes on a mono-block mesh using a conventional staggered-grid approach. All field variables except velocities are stored at the grid nodes, while the velocities themselves are stored at staggered cell-face locations which lie between the nodes. The finite-volume equations for each variable are derived by integrating the partial differential equations over each control volume. Fully implicit backward differencing is employed for the transient terms, and central differencing is used for the diffusion terms. The convective terms are discretized using hybrid (central or upwind) differencing.
The integration procedure results in a coupled set of algebraic finite-volume equations that express the value of a variable at a grid node in terms of the values at neighboring grid points and the nodal value at the old-time level. The finite-volume equations are solved iteratively using SIMPLEST algorithm of Spalding, which is embodied in PHOENICS for the solution of single-phase. The algorithm is a segregated solution method that employs pressure-velocity coupling to enforce mass conservation by solving a pressure-correction equation and making corrections to the pressure and velocity fields.

\begin{table*}[htbp]
\scriptsize
  \centering
  \caption{Discretization errors on four non-uniform grids. The parameters are $Bd=0$, $Bi=0.2$, $B_\mathrm{ev}=0$, and $Ma=200$.} \label{tabmesh}
\setlength{\leftskip}{-30pt}
 
 \begin{tabular}{cccccccc}
   \toprule
   Grid No.   &   $N_r \times N_\phi \times N_z$ (CVs)     &   $\theta_\mathrm{min}$    & $\epsilon_\theta$ (in \%)    &   $u_\mathrm{max}$  & $\epsilon_u$ (in \%)  &Kinetic Energy($E$)  &  $\epsilon_E$ (in \%)  \\
    \midrule
  1  &  $100\times20\times20$      &   0.8338  & 0.25 & 1.707 $\times$ 10$^{-2}$   & 5.73  & 6.106 $\times$ 10$^{-4}$  & 4.19   \\ 
  2  &  $150\times30\times30$      &   0.8326  & 0.11 & 1.765 $\times$ 10$^{-2}$   & 2.52  & 5.974 $\times$ 10$^{-4}$  & 1.94   \\
  3  &  $200\times40\times40$      &   0.8321  & 0.05 & 1.793 $\times$ 10$^{-2}$   & 0.98  & 5.875 $\times$ 10$^{-4}$  & 0.25   \\
  4  &  $250\times50\times50$      &   0.8317  &    - & 1.811 $\times$ 10$^{-2}$   & -   & 5.860 $\times$ 10$^{-4}$  &     -     \\
     \bottomrule
    \end{tabular}
\end{table*}

\subsection{{\color{black}Model} validation}
We begin with the estimation of discretization errors. We performed computations on four non-uniform grids consisting of $N_r \times N_\phi \times N_z$ control volumes (CVs), where $N_r$, $N_\phi$, and $N_z$  denote the number of control volumes in the radial, azimuthal, and axial direction, respectively. The fine grid is near the walls and the free surface whereas the coarse grid is in the middle of the solution domain. Taking the finest grid $250\times50\times50$ as an estimated grid-independent solution, we compute the relative difference in the solutions $\epsilon$ (in \%) as compared to the solution obtained on the finest grid, for the minimum temperature of the liquid layer ($\theta_\mathrm{min}$), the maximum velocity ($u_\mathrm{max}$), 
and the total momentum ($E$). As shown in Table \ref{tabmesh}, these differences are less than 1\% on the $200\times40\times40$ CV grid. Therefore all the subsequent computations are performed on this grid.

{\color{black}Before presenting numerical results there is a need to validate the present numerical model by comparing simulation results with previously reported experimental data. To that end, the numerical code has been accommodated to simulate an experimental study~\cite{zhu2010coupling}. The experiment consists of an evaporating thin liquid layer ($h = 2$ mm) of 0.65 cSt silicone oil ($Pr=6.7$) subjected to a lateral temperature difference $\Delta T=$ \SI{2}{\degreeCelsius}.  We used a combined Biot number equal to 0.2 (i.e., $Bi + B_\mathrm{ev}=0.2$), which is estimated from the experimental data. The surface temperature profile obtained from the numerical simulation with these parameters is compared with the experimental measurement by a thermocouple. The comparison in Fig.~\ref{figa1} shows a very good agreement. The obtained steady-state unicellular flow pattern also agrees with the PIV (Particle Image Velocimetry) result (not shown).  We also provide an additional comparison with experimental observations in Section~\ref{sec:4-3}, where we demonstrate the present numerical model includes the essential ingredients capable of describing the emergence of pattern formation in an evaporating thin layer of FC-72.

\begin{figure}[htbp]
	\centering
	\includegraphics[width=0.7\textwidth]{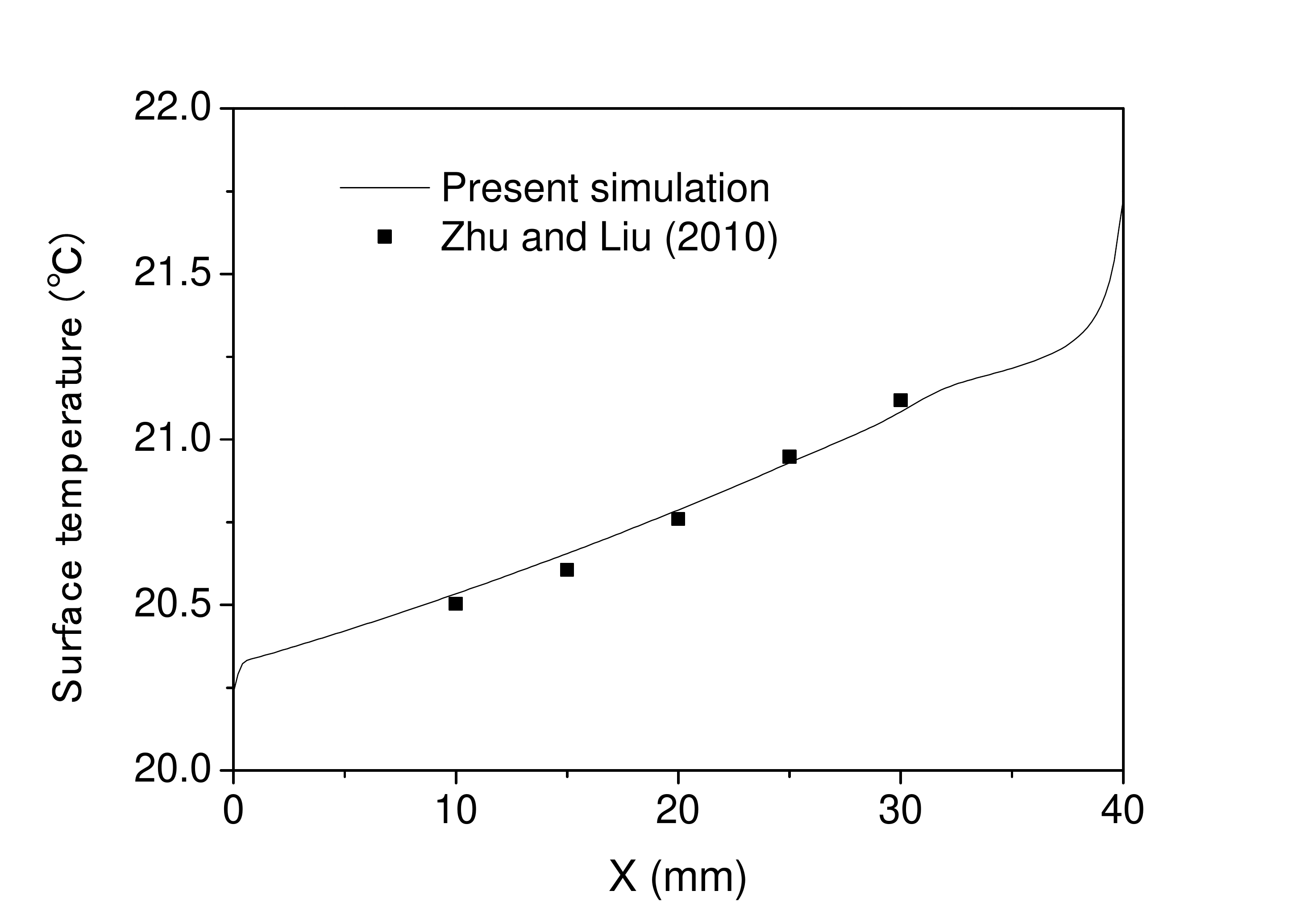}
	\caption{{\color{black}Comparison of the surface temperature profile between the present numerical simulation and the experimental measurement by Zhu and Liu \cite{zhu2010coupling}}}\label{figa1}
\end{figure}
}
\section{Results and Discussion}\label{Sec:Results}

\subsection{Steady axisymmetric flow}
\subsubsection{Thermocapillary-driven flow}
The axisymmetry of the geometry and steady axisymmetric boundary conditions allow for a steady axisymmetric basic flow. It is in particular true for relatively small Marangoni numbers. As an example, we use the setting of $Bd=0$, $Bi=0.2$, $B_\mathrm{ev}=0$ and $Ma=300$ to illustrate the characteristic features of the purely thermocapillary flow in the absence of evaporation.
\begin{figure}[htbp]
	\centering
	\includegraphics[width=0.6\textwidth]{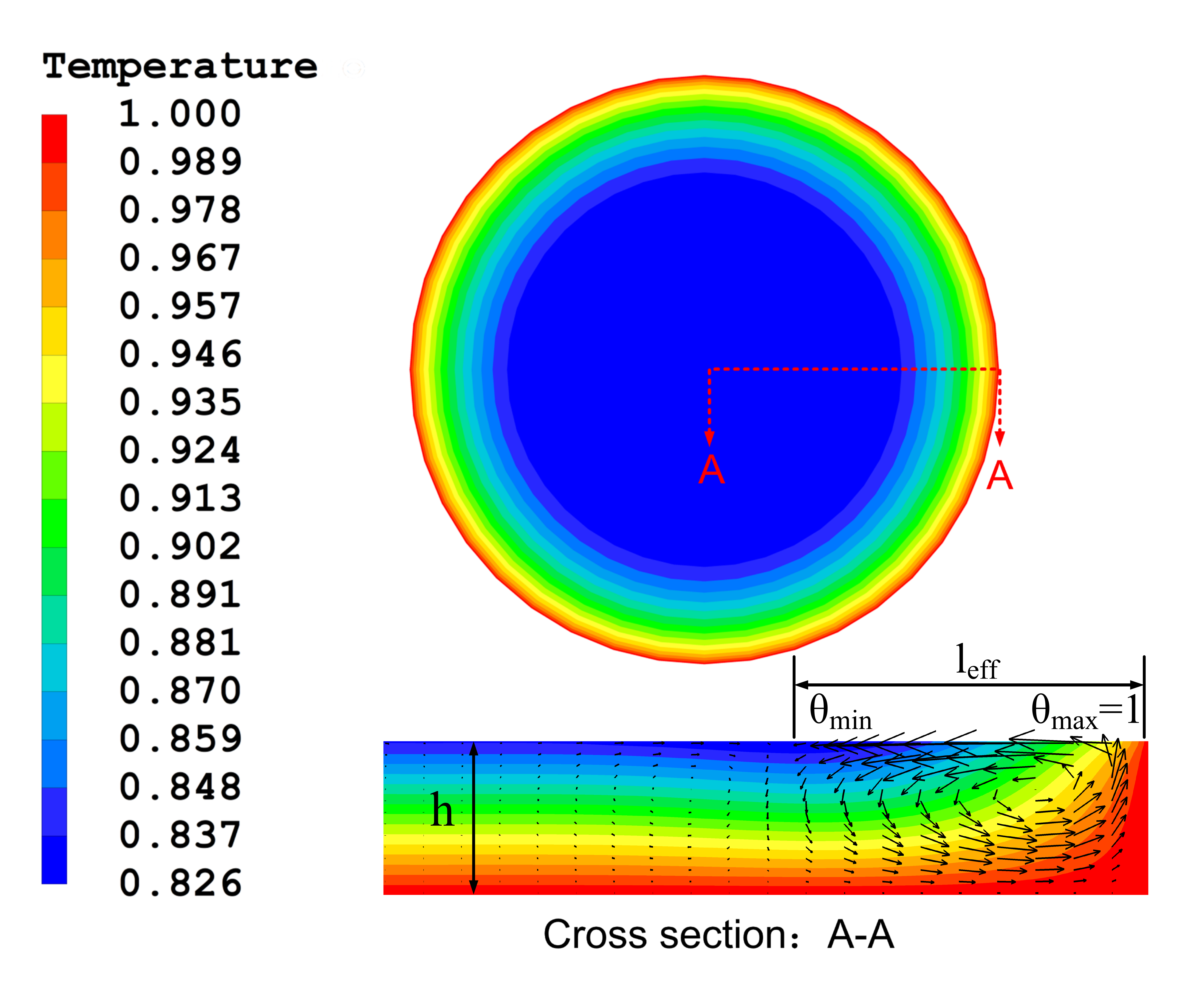}
	\caption{Surface temperature field (top), temperature and velocity fields (bottom) on the cross-section. The parameters are $Bd=0$, $Bi=0.2$, $B_\mathrm{ev}=0$, and $Ma=300$. The effective driving temperature difference is indicated as ($1 - \theta_\mathrm{min}$) over an effective distance $l_\mathrm{eff}$.} \label{fig3}
\end{figure}

Due to the presence of the heated sidewall, which is at constant temperature ($\theta =1$), a positive temperature gradient arises close to the sidewall. The temperature gradient induces locally high thermocapillary stresses, which give rise to a fluid motion from the hot (periphery) to the cold (middle) region on the free surface. The flow penetrates the bulk through viscous coupling to the motion on the free surface, forming a counter-rotating cell  in the layer near the sidewall, as shown in Fig.~\ref{fig3}. It is also shown that the central region of the layer is almost motionless. The center of the vortex appears to move toward the center of the layer when the Marangoni number is further increased. This thermocapillary flow structure is similar to the flow occurring in an open rectangular cavity whose vertical endwalls are maintained at different temperatures (e.g.,~\cite{Zebib_1985,mercier2002influence}).
\begin{figure}[htbp]
	\centering
	\includegraphics[width=0.6\textwidth]{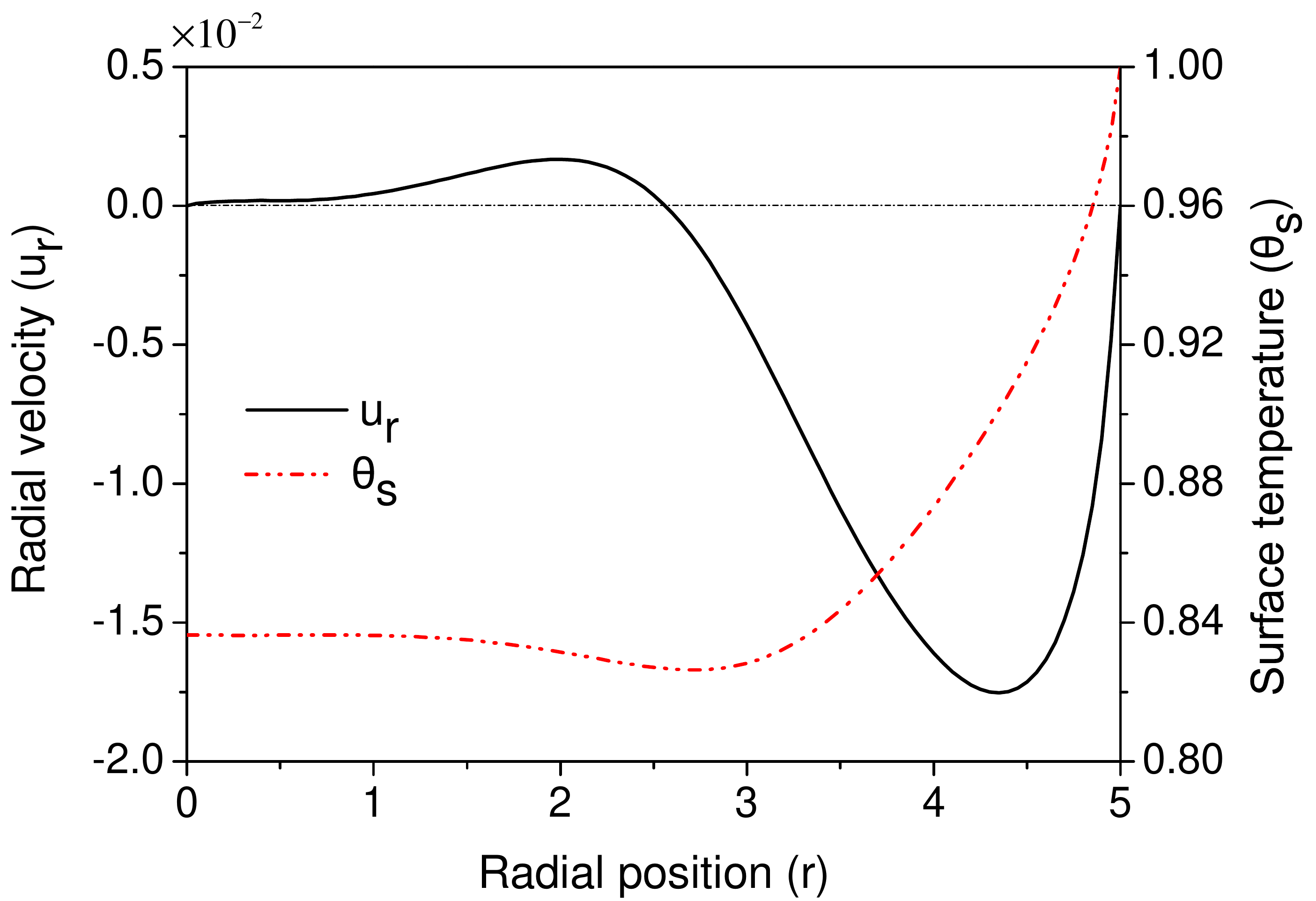}
	\caption{The radial velocity ($u_r$) profile (solid line, left axis) and the temperature (dash-dotted line, right axis) distribution on the free surface. The parameters are the same as in Fig.~\ref{fig3}.}\label{fig4}
\end{figure}

The flow is driven by thermocapillarity, so we define an effective driving temperature difference as $1-\theta_\mathrm{min}$, with $\theta_\mathrm{min}$ a minimum temperature on the free surface ($\theta_\mathrm{max}=1$). The coupling between the temperature gradient on the free surface and the surface flow is clearly shown in Fig.~\ref{fig4}. It is seen that the surface temperature decreases rapidly, reaching a minimum temperature $\theta_\mathrm{min}$ where the effective length $l_\mathrm{eff}$ ($\approx 2.5$) is located. These large temperature gradients induce relative large radially inward surface flow in the region near the sidewall.  The surface temperature distribution is then followed by a plateau with a weak (local) maximum developing at $r \approx 2$ which reduces the driving forces over this large portion of the surface; the outward surface flow becomes weak and vanishingly small near the axis. The flow reversion is due to a change of sign in the surface temperature gradient, at the point $\theta = \theta_\mathrm{min}$.

\subsubsection{Influence of the Biot number}

An important influential parameter on the thermocapillary flow is the Biot number $Bi$, which characterizes the heat transfer in the ambient gas. An adiabatic free surface, i.e., $Bi=0$, is not allowed since the temperature would be uniform in the whole domain corresponding to the imposed wall temperature $\theta =1$ and, therefore there would be no fluid motion. On the contrary, for very large values of $Bi$ the surface temperature tends to be imposed by the ambient gas, i.e., $\theta_s \to 0$. Hence, increasing the Biot number reduces the interfacial temperature and increases the temperature gradients on the free surface, which enhances the thermocapillary flow. 

\begin{figure}[htbp]
	\centering
	\includegraphics[width=0.6\textwidth]{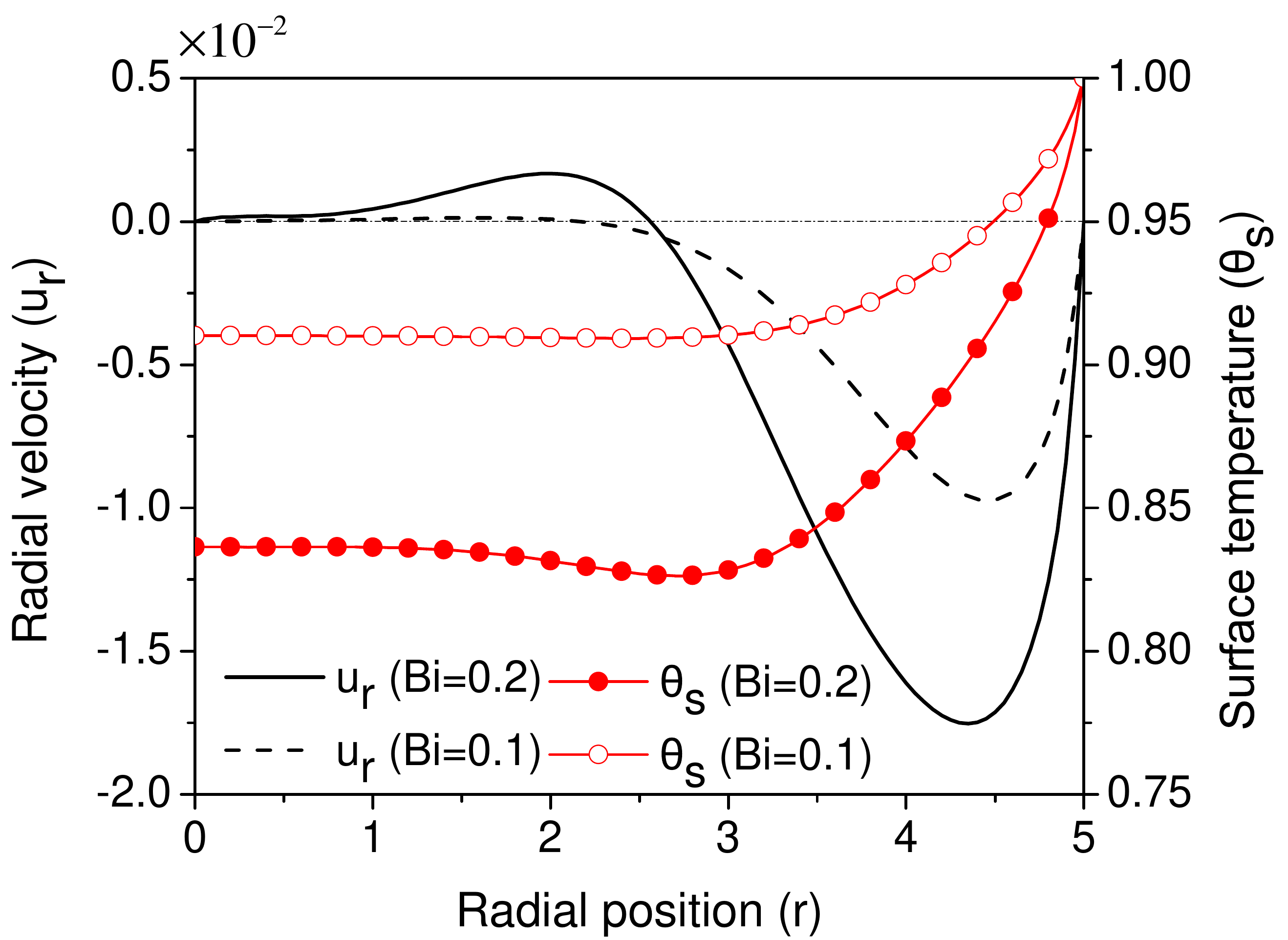}
	\caption{The radial velocity profile and the temperature distribution on the free surface for two Biot numbers, $Bi=0.1$ and $Bi=0.2$, for $Bd=0$ and $Ma=300$.}\label{fig5}
\end{figure}

Figure~\ref{fig5} shows the temperature distribution along the free surface and the radial velocity profile for two Biot numbers. Even a slight increase in the Biot number, i.e., from 0.1 to 0.2, can result in a notable decrease in the surface temperature and an increase of the surface flow. Again, the radial velocity profile is dictated by the temperature distribution on the free surface, as discussed before.

\subsubsection{Evaporative cooling effect} 
Owing to the high volatility of FC-72 liquid, the evaporative cooling, which is described by a high evaporative Biot number $B_\mathrm{ev}$ (=2), has significant effects on the thermocapillary flow. As can be seen clearly in Fig.~\ref{fig6}, the latent heat of evaporation substantially reduces the interfacial temperature with a tendency to make the temperature closer to the saturation temperature, much higher temperature gradients are created near the sidewall compared to the situation without evaporation, which greatly enhances the flow. The enhanced surface flow near the sidewall amplifies the thermocapillary-driven vortex with its center shifting inward. The flow inversion on the free surface occurs at $\theta = \theta_\mathrm{min}$. The reversed flow is, however, very weak due to the presence of a  temperature plateau; the effect of evaporation is insignificant in this region except for shrinking the plateau.

\begin{figure}[htbp]
	\centering
	\includegraphics[width=0.6\textwidth]{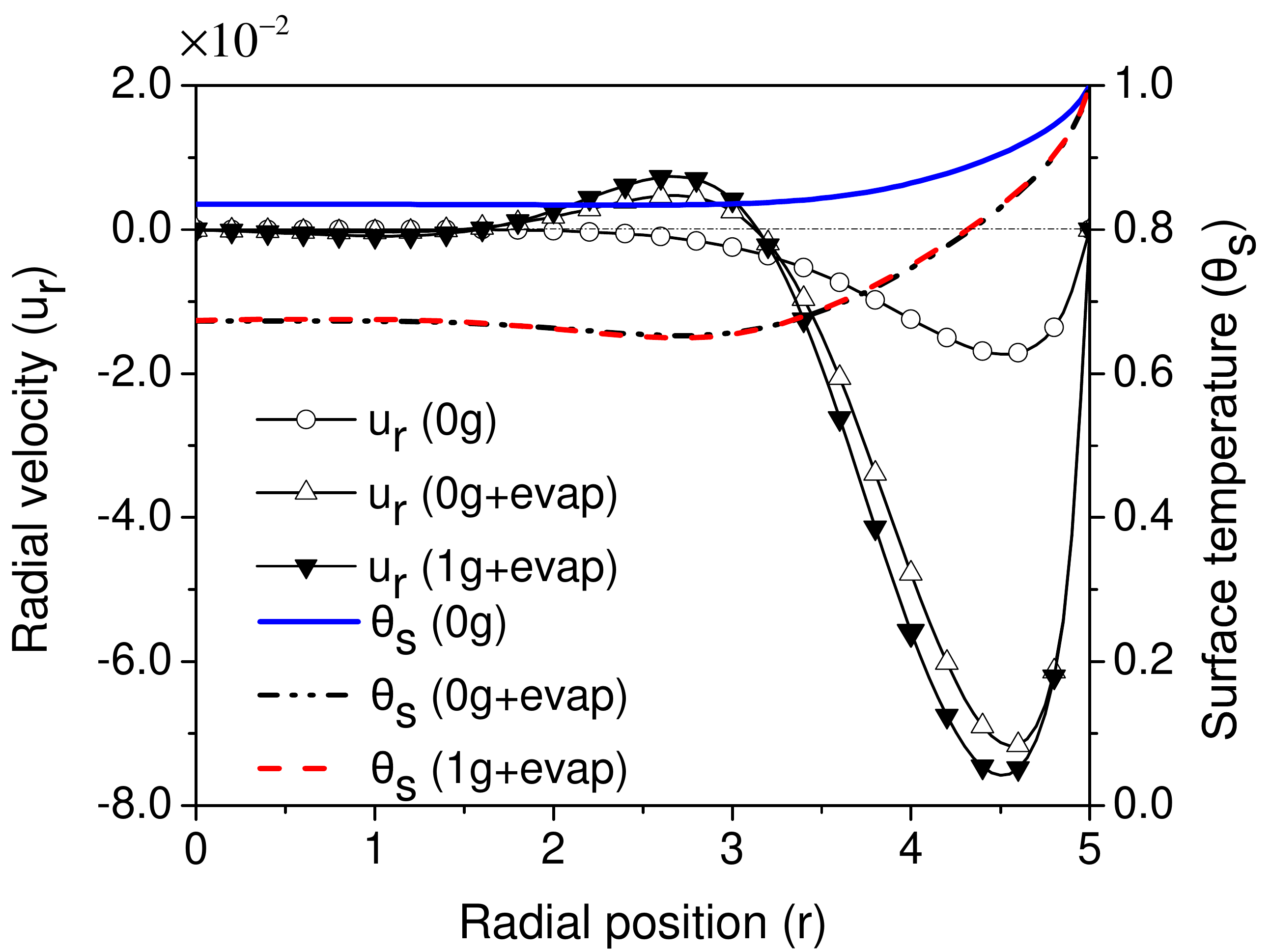}
	\caption{A comparison of the interfacial temperature distributions and radial velocity profiles with and without evaporation, both for 0g and 1g, for $Bi=0.2$, and $Ma=100$.} \label{fig6}
\end{figure}

\subsubsection{Thermocapillary-buoyant flow} \label{sub:gravity}
The combined thermocapillarity, buoyancy, and evaporation are also shown in Fig.~\ref{fig6}. In the present setting, the buoyancy-driven flow has little effect on the temperature distribution on the free surface. Buoyancy forces slightly augment the thermocapillary flow both in the radially inward flow (close to the sidewall) and in the outward flow (in the central region), so the flow is predominantly driven by thermocapillary forces, slightly altered by buoyancy effects.

In normal gravity with a dynamic Bond number of $Bd=1.2$, one would expect a pronounced effect of buoyant body forces on the thermocapillary-driven flow. The reason for such a small effect is the reduced temperature difference across the layer between the bottom wall ($\theta_w$) and the upper surface ($\theta_s$). Since $\theta_w - \theta_s < 0.4$ (cf. Fig.~\ref{fig6}), the effective dynamic Bond number is de facto reduced by at least 60\%, leading to $Bd_\mathrm{eff} \approx 0.5$.

To close this subsection, we should mention that the flow characteristics are very much dependent on the Marangoni number and, therefore all the above description is limited to the axisymmetric flow, i.e.,  a low-Marangoni-number flow.

\subsection{Flow transition}
\subsubsection{Prediction of the transition}

By increasing the Marangoni number, the numerical simulations reveal the emergence of a new steady-state flow; the axisymmetric flow undergoes a symmetry breaking, giving rise to a fully three-dimensional (3D) steady flow. To predict a critical threshold of stability (i.e., the critical Marangoni number $Ma_c$) beyond which the transition takes place, we calculate the ratio of $E_\phi/E_0$, with $E_\phi$ ($\equiv \int u^2_\phi \,\mathrm{dV}$) the azimuthal kinetic energy and $E_0$ ($\equiv \int (u^2_r + u^2_\phi +u^2_z) \, \mathrm{dV}$) the total kinetic energy of the flow. When this ratio is less than a certain percent, say 1\%, the computed flow is considered axisymmetric, otherwise, it is within the 3D flow regime. Figure~\ref{fig7} gives an example of how $Ma_c$ is determined as a function of the ratio $E_\phi/E_0$. In this setting, the critical Marangoni number $Ma_c$ is predicted to be about 340. While we emphasize the steady-state nature of the 3D flow, we haven't attempted to determine the nature of the bifurcation -- namely whether it is supercritical or subcritical. Such an attempt would involve considerable numerical computations of the flow very close to the transition point, which is certainly beyond the scope of the present work. However, a linear stability analysis (e.g.,~\cite{Xun_2008,Xun_2009}) of the basic state can be applied to precisely determine the instability threshold and identify the underlying mechanisms.

\begin{figure}[htbp]
	\centering
	\includegraphics[width=0.6\textwidth]{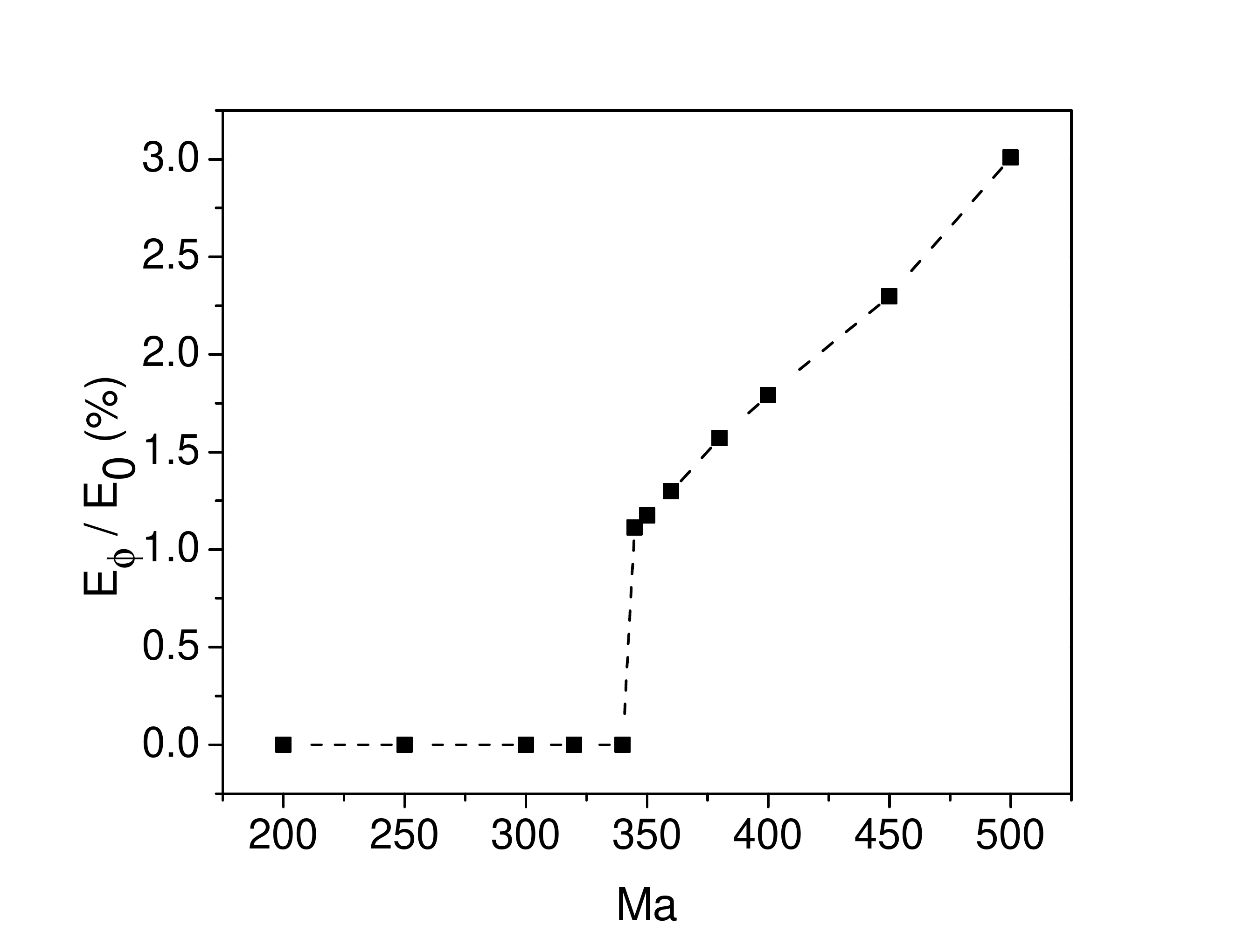}
	\caption{The ratio $E_\phi/E_0$ as a function of the Marangoni number for $Bd=0$, $Bi=0.2$, and $B_\mathrm{ev}=0$.} \label{fig7}
\end{figure}

\subsubsection{Critical Marangoni number}
A systematic parametric study is carried out to investigate the effects of influential parameters on the critical Marangoni number.
As shown before, one of the most important parameters is the Biot number. Therefore, we plot in Fig.~\ref{fig8} the variation of the critical Marangoni number $Ma_c$ vs. the Biot number $Bi$ under different conditions. The Biot number $Bi$ is assumed to be relatively small, ranging from 0.05 to 1, while the evaporative Biot number $B_\mathrm{ev}$ is taken to be 2, based on our ground and space experiment.

\begin{figure}[htbp]
	\centering
	\includegraphics[width=0.6\textwidth]{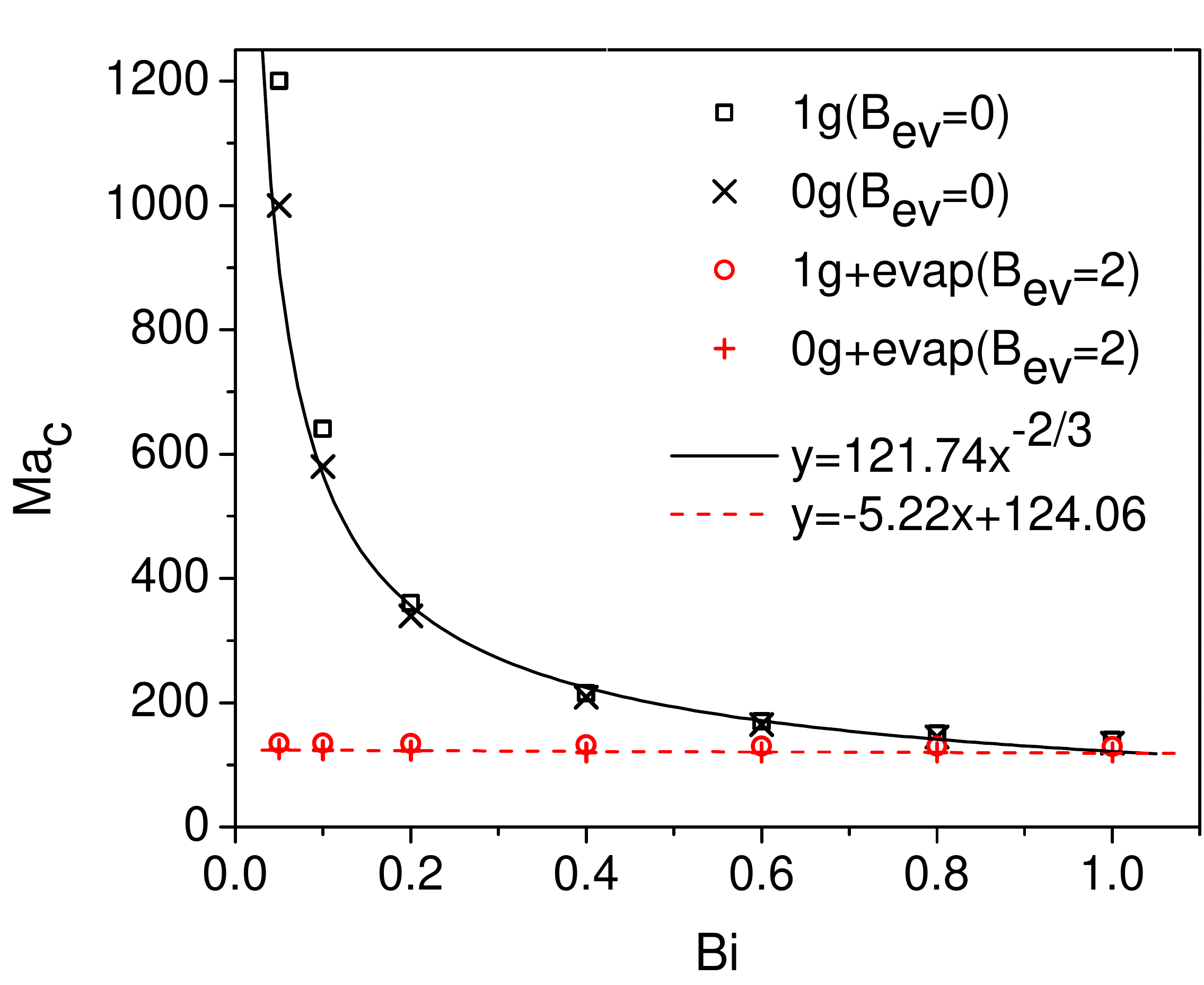}
	\caption{Critical Marangoni number $Ma_c$ as a function of the Biot number $Bi$ for 0g and 1g, and without evaporation (black) and with evaporation (red), along with the best-fit curves.} \label{fig8}
\end{figure}

It is clearly shown that the critical Marangoni number depends strongly on the Biot number in the absence of evaporation since $Ma_c$ remarkably follows a power law, i.e., $Ma_c \sim Bi^{-2/3}$ ($Bi \le 1$). For a larger Biot number, say $Bi > 1$, a saturation regime is reached wherein the critical Marangoni number remains roughly constant ($Ma_c \approx 120$). This explains why the critical Marangoni number varies slightly with $Bi$ in the case of evaporation;  the evaporative cooling results in a combined Biot number greater than 2.

In dynamic thermocapillary liquid layers, viscous and thermal diffusion has a stabilizing effect by providing restoring forces~\cite{smith1983instabilities,schatz2001experiments}. The numerical results shown in Fig.~\ref{fig8} give some insight into the mechanisms of instability.  The enhanced thermocapillary-driven flow due to evaporation greatly contributes to the destabilization, leading to a lower bound of the critical Marangoni number approximating 120. Counter-intuitively, buoyancy effects stabilize the basic flow, especially for very small values of $Bi$ (i.e., $Bi <0.1$). For larger values of $Bi$, the effects of buoyancy are virtually invisible (cf. Fig.~\ref{fig8}).

\subsubsection{Effective critical Marangoni number} 

We introduce the effective Marangoni number based on an effective thermocapillary-driven velocity $u_\mathrm{eff}$,
\begin{equation}
 Ma_\mathrm{eff}=\frac{u_\mathrm{eff}h}{ \alpha}, \label{eq:Maeff}
\end{equation}
with $u_\mathrm{eff}$ being given by a balance of stresses in the radial direction on the free surface (Eq.~\ref{eq:dym}),
\begin{equation}
 u_\mathrm{eff}=\frac{\gamma \Delta T \Delta \theta }{\mu l_\mathrm{eff}},
\end{equation}
where $\Delta \theta$ ($\equiv 1 - \theta_\mathrm{min}$) is the effective driving temperature difference over the distance $l_\mathrm{eff}$ (cf. Fig.~\ref{fig3}).
We then arrive at a relationship between the effective critical Marangoni number $Ma^c_\mathrm{eff}$ and the critical Marangoni number $Ma_c$,
\begin{equation}
 Ma^c_\mathrm{eff}=\frac{\Delta \theta}{l_\mathrm{eff}}Ma_c.  
\end{equation}
We must point out however that $Ma_\mathrm{eff}$ is not a controlled parameter, unlike $Ma$. Introducing it aims to gain a better understanding of the role of evaporation in the instability mechanisms.

Figure~\ref{fig9} shows the variation of $Ma^c_\mathrm{eff}$ as a function of $Bi$ under different conditions. In contrast to Fig.~\ref{fig8}, the effective critical Marangoni number now increases with the Biot number. More importantly, $Ma^c_\mathrm{eff}$ under evaporation is significantly larger than its counterpart without evaporation. These results suggest the latent heat of evaporation also plays a stabilizing role. The stabilizing behavior is attributed to the reduced interfacial temperature in the region where the temperature exhibits a plateau (cf. Fig.~\ref{fig6}). Hence, evaporation plays a twofold role: on the one hand, the evaporative cooling creates locally higher radial interfacial temperature gradient (near the sidewall) which enhances the thermocapillary-driven flow -- evaporation destabilizes the flow; on the other hand, evaporation stabilizes it by removing heat from the interface, especially in the central part of the liquid layer. Such an interplay has already been pointed out in~\cite{Karapetsas_2012} for evaporating sessile drops deposited on a heated surface, though the underlying mechanisms are not the same. 

\begin{figure}[htbp]
	\centering
	\includegraphics[width=0.6\textwidth]{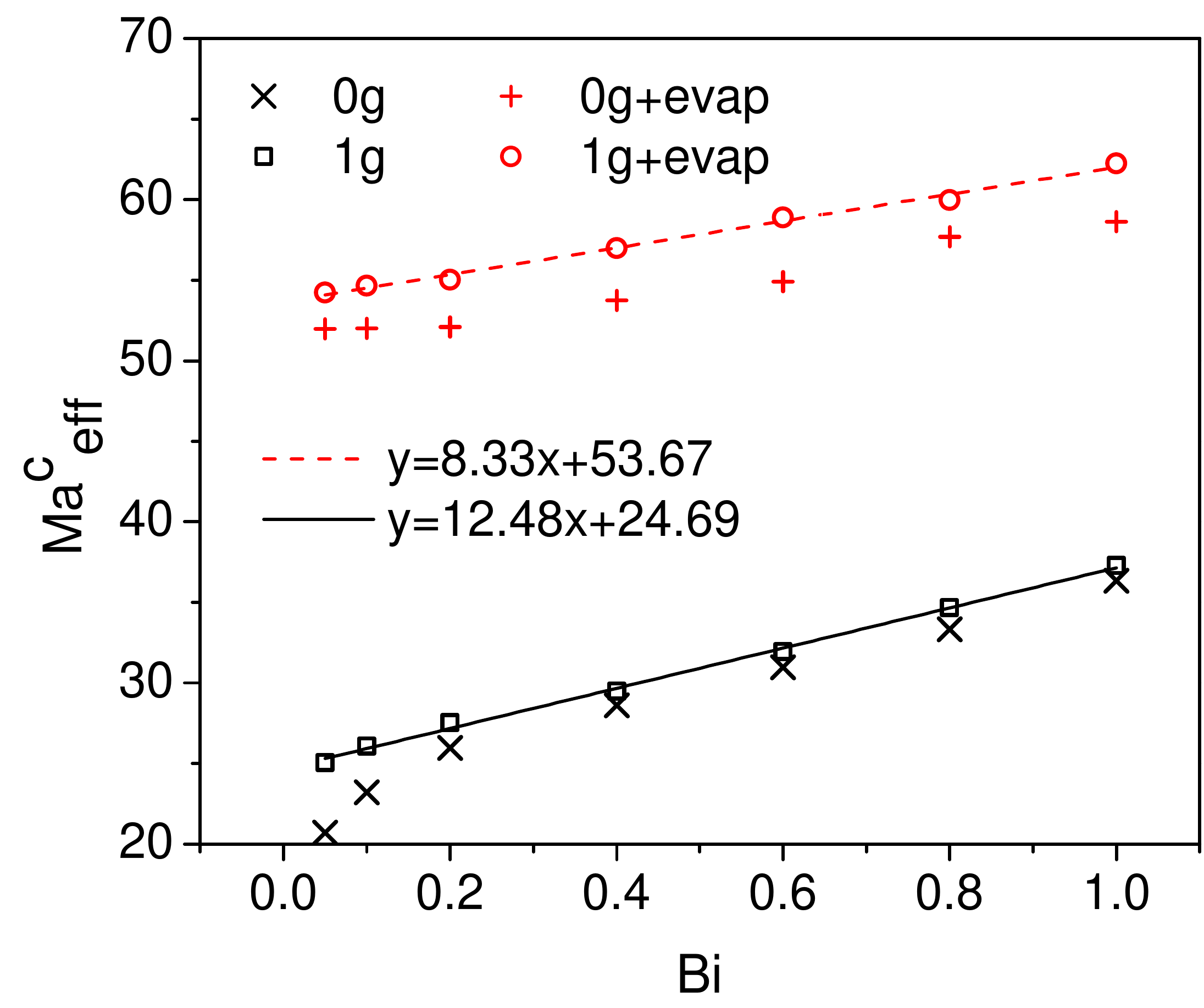}
	\caption{Effective critical Marangoni number $Ma^c_\mathrm{eff}$ as a function of the Biot number $Bi$ for 0g and 1g, and without evaporation (black) and with evaporation (red), along with the best-fit curves.} \label{fig9}
\end{figure}

Interestingly, the effective critical Marangoni number scales almost linearly with the Biot number when $Bi>0.2$.
This can be traced back to the ratio $\Delta \theta/l_\mathrm{eff}$, the effective driving temperature gradient, which is nearly an increasing linear function of the Biot number (Fig.~\ref{fig10}), while $Ma_c$ remains almost unchanged (Fig.~\ref{fig8}). Also, Fig.~\ref{fig9} further confirms the stabilizing influence of buoyancy forces, which now can be explained with a smaller $\Delta \theta/l_\mathrm{eff}$ due to buoyancy effects, as shown in Fig.~\ref{fig10}. 

{\color{black} We have also performed simulations with liquid-layer heights $h$ varying from 2  to 4 mm to examine the gravity (buoyancy) effect. Increasing the liquid height, namely increasing buoyancy effect, leads to augment the surface flow not only in the radially inward flow (close to the sidewall) but also in the outward flow (in the central region). As a result, the flow transition is not so much affected since $Ma_c=140$ for $h=4$ mm,  compared to $Ma_c=134$ for $h=2$ mm.  The results also show that buoyancy in the liquid layer has a stabilizing effect. 
}


\begin{figure}[htbp]
	\centering
	\includegraphics[width=0.6\textwidth]{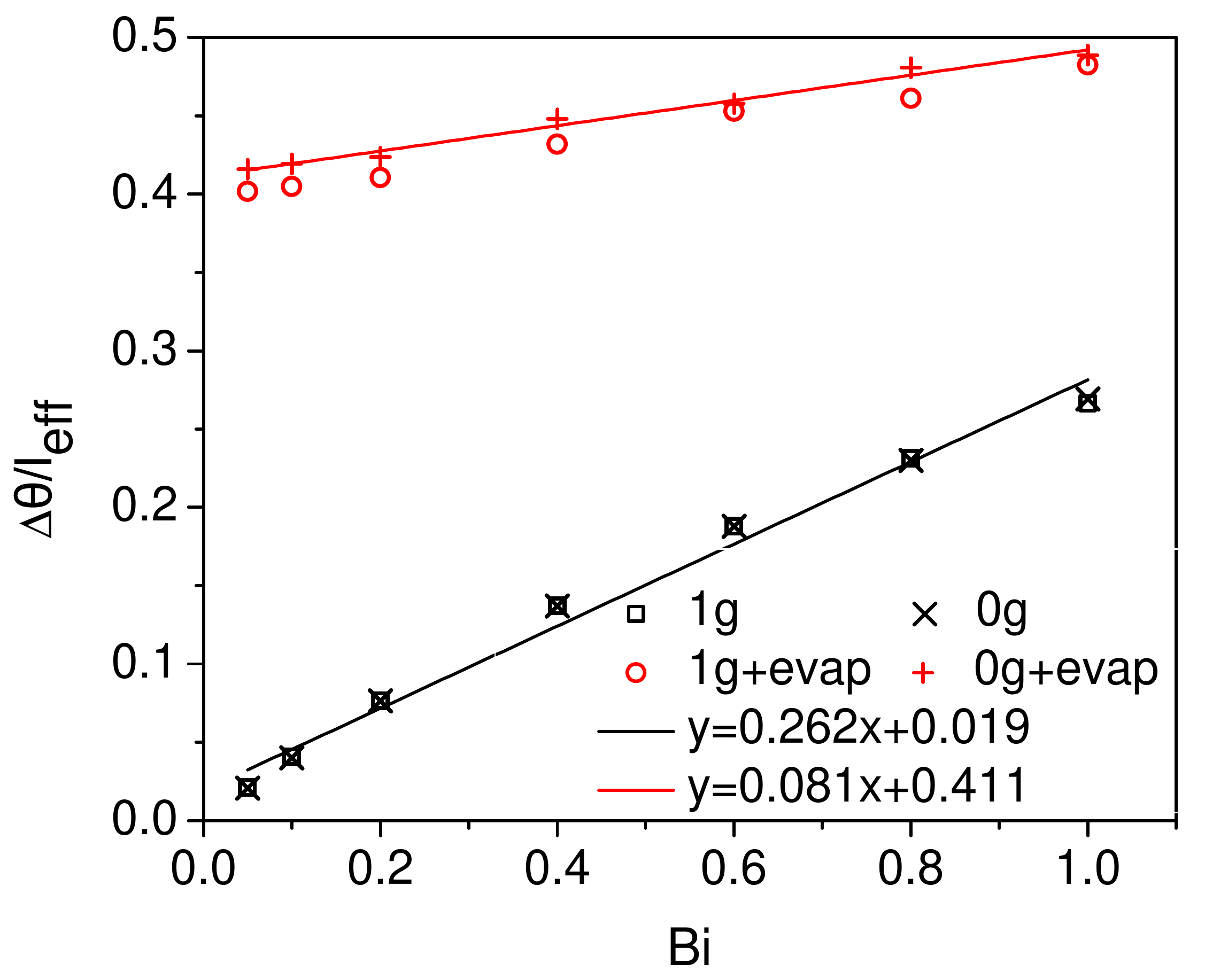}
	\caption{Variation of $\Delta \theta/l_\mathrm{eff}$ as a function of the Biot number for different conditions under study, along with the best-fit curves.} \label{fig10}
\end{figure}

\subsection{Beyond the transition} \label{sec:4-3}

Further increasing the Marangoni number beyond its critical threshold leads to symmetry-breaking which gives rise to fully 3D, steady-state, convective flows. 
Figure~\ref{fig11} illustrates such a flow at $Ma=500$, far beyond the critical point, i.e., $Ma_c\approx340$. In contrast to Fig.~\ref{fig3}, thermocapillary-driven flow now exhibits drastic changes in the vortex structures; stable multiple vortex patterns form in the layer with flow crossing through the central axis ($r=0$).

{\color{black}
To illustrate a typical pattern formation developed in an evaporating thin layer of FC-72 at high Marangoni numbers, we performed two 1g-simulations, one at $Ma=17465$ and $Bd=1.2$, which correspond to a real ground FC-72 experiment ($h=2$ mm) under an applied temperature difference around \SI{2}{\degreeCelsius}, and the other at $Ma=6113$ and $Bd=0.15$, corresponding to the experimental conditions  conducted at $h=0.7$ mm and $\Delta T\approx$ \SI{2}{\degreeCelsius}. A comparison of the numerical predictions with the images produced using infrared thermography is shown in Fig.~\ref{figa2}. The numerical simulations revealed the spontaneous formation of various patterns, depending mainly on the depth of the liquid layer. As shown in Fig.~\ref{figa2}, a thinner evaporating layer exhibits smaller-scale thermal patterns on the surface, in qualitative agreement with the experimental observations. This favorable comparison demonstrates that the present numerical model contains the physical mechanisms that drive the development of pattern formation in an evaporating volatile liquid layer.}

{\color{black}
At these high Marangoni numbers, the question arises as to whether the flow remains laminar flow behavior as has been assumed in the present work. To estimate the critical Reynolds number $Re_c$ for the transition of laminar steady-state flow to oscillatory one, which is still far from the turbulent flow, we make use of the reported results in the literature (e.g., Refs.~\cite{sim2002effect,kamotani2000microgravity}). It turns out that the critical Reynolds number $Re_c$ lies in the range of 1500--4500 for the onset of oscillatory flow. A Marangoni number of 17465 for a FC-72 liquid ($Pr=12.34$) leads to $Re=Ma/Pr\approx 1415$, thereby justifying the assumption of laminar flow.
}

\begin{figure}[htbp]
	\centering
	\includegraphics[width=0.6\textwidth]{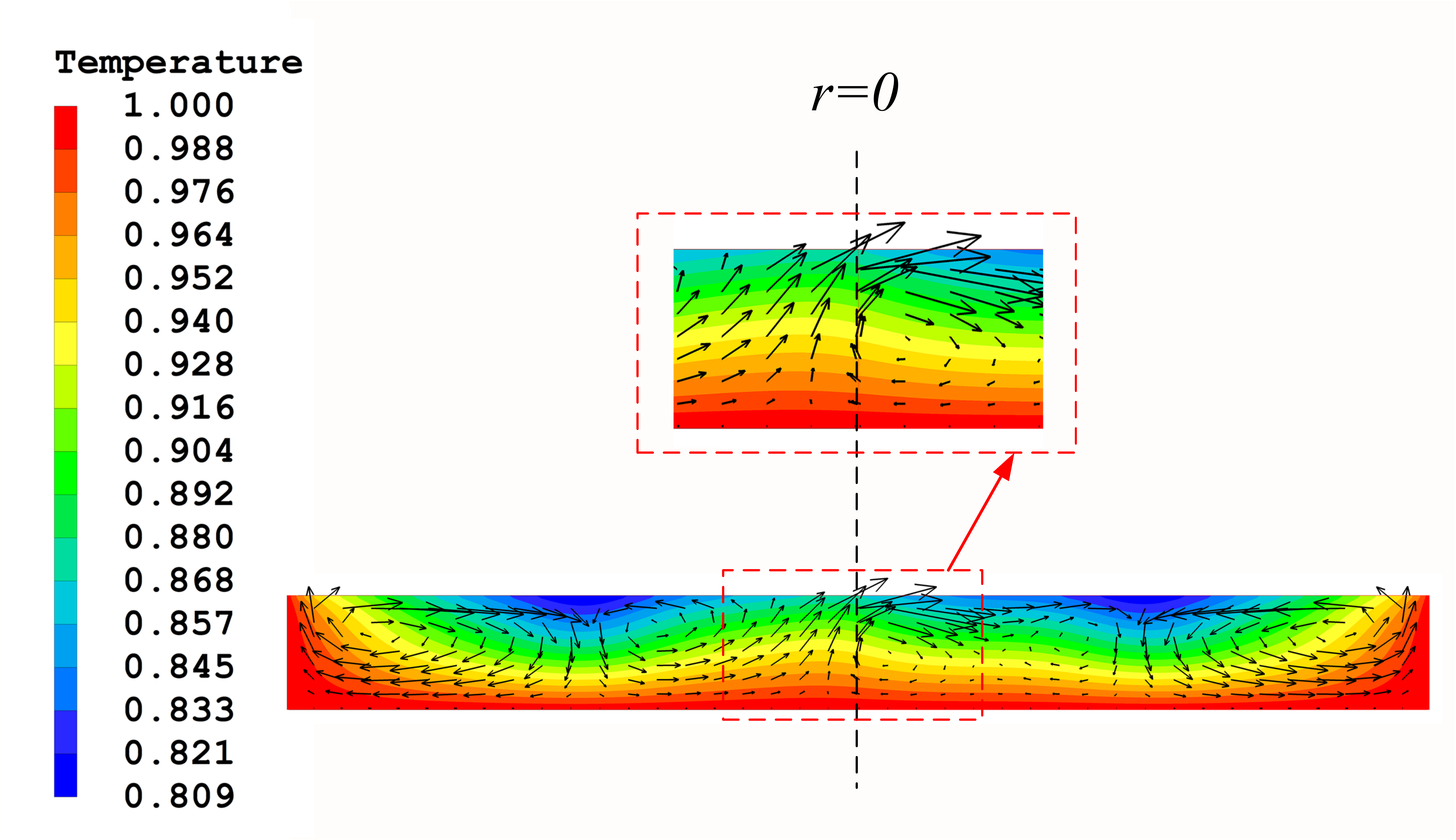}
	\caption{Temperature and velocity fields on a cross section in a fully 3D thermocapillary flow. The parameters are $Bi=0.2$, $Bd=0$, $B_\mathrm{ev}=0$, and $Ma=500$.} \label{fig11}
\end{figure}

\begin{figure}[htbp]
	\centering
	\includegraphics[width=0.6\textwidth]{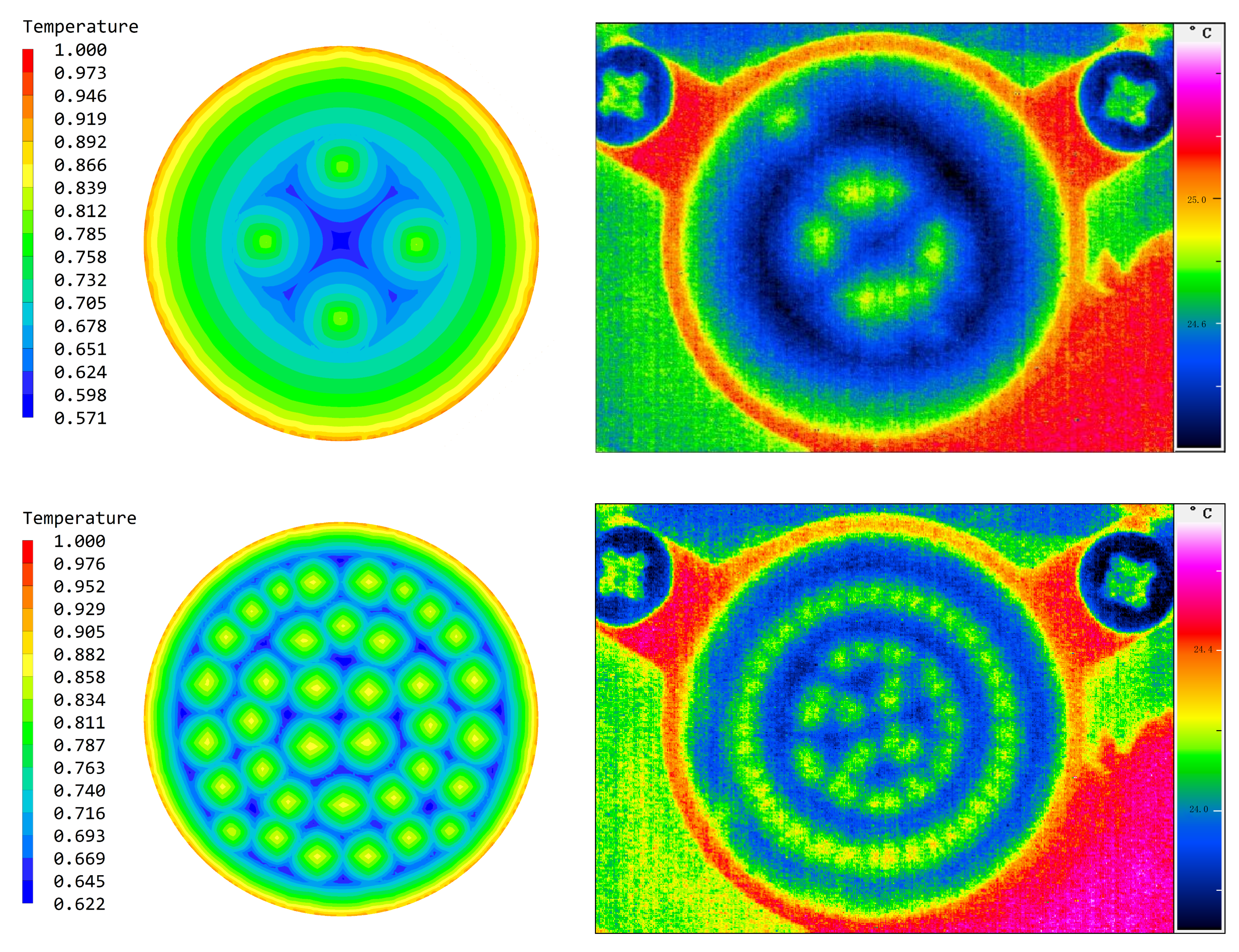}
	\caption{{\color{black}A comparison of the surface temperature field obtained between from numerical simulations (left panel) and using  infrared thermography in ground experiments with FC-72 (right panel). The parameters of simulation are (top): $Bi=0.2$, $Bd=1.2$, $B_\mathrm{ev}=2.0$, and $Ma=17465$, corresponding to the experimental conditions conducted with $h=2$ mm and $\Delta T\approx$ \SI{2}{\degreeCelsius}; (bottom): $Bi=0.2$, $Bd=0.15$, $B_\mathrm{ev}=2.0$, and $Ma=6113$, corresponding to the experimental conditions conducted with $h=0.7$ mm and $\Delta T\approx$ \SI{2}{\degreeCelsius}.}} \label{figa2}
\end{figure}

Given that the present numerical simulation is based on a one-sided model of evaporation which ignores the dynamics of the gas phase, the question as to whether such kind of 3D flows is stable and when and how it eventually becomes a time-dependent flow (e.g., in the form of hydrothermal waves propagating azimuthally in the pool) is not addressed. Answering this question remains a computationally challenging task. Furthermore, if we attempt to make any quantitative comparison with experimental observations it would also be necessary to consider a more elaborate two-phase model that accounts for heat, mass, and momentum transport in the gas~\cite{Chen_2017a}. Indeed, in their study of the convective instability of the liquid layer subject to a horizontal temperature gradient, Grigoriev and Qin \cite{grigoriev2018effect} pointed out the shortcomings of the one-sided model and advocated accounting for the transport of heat and vapor through the gas phase, particularly when the applied temperature gradient becomes larger.

Finally, the question regarding whether the predicted transition points (i.e., $Ma_c$) are detectable in an experiment needs to address. The answer is somewhat disappointing. The prediction of the critical Marangoni numbers, presented in Fig.~\ref{fig8}, requires a minute applied temperature difference $\Delta T$, on the order of \SI{0.1}{\degreeCelsius}  at the most, for the thermocapillary-driven flow to be axisymmetric. So small temperature differences are difficult or even impossible to realize in practice. This is why we didn't observe any axisymmetric flows in our experiments (with FC-72); the applied temperature differences are far beyond the predicted ones. With a much less volatile fluid, for instance, 5cSt silicon oil ($Pr=68$), our numerical model, however, predicts a feasible temperature difference, i.e.,  $\Delta T \approx$ \SI{1}{\degreeCelsius}.

\section{Summary and concluding remarks}\label{Sec:Conclusion}

We have carried out a numerical study on an evaporating liquid layer (FC-72) enclosed in a cylindrical cell and surrounded by a passive gas. The numerical simulations are based on an enhanced one-sided model of evaporation that ignores the dynamics of the gas phase but includes the effect of conductive heat transport through the gas phase and the effect of the latent heat associated with a phase change at the interface. These effects are described by the heat transfer Biot number $Bi$ and the evaporative Biot number $B_\mathrm{ev}$, respectively. Calculations have been performed to elucidate the flow features and predict the flow transition from axisymmetric to fully 3D patterns. 

For a sufficiently small Marangoni number, the flow remains steady and axisymmetric. It can be characterized by a primary vortex near the sidewall, which is predominantly driven by thermocapillarity and slightly modified by buoyancy effects (in 1g). Both the conductive heat transport through the gas phase and the latent heat of evaporation reduce the interfacial temperature and increase the thermal gradient on the free surface near the sidewall, resulting in locally higher thermocapillary stresses. Due to a larger value of $B_\mathrm{ev}$ relative to $Bi$, the evaporative cooling effect significantly enhances the thermocapillary flow.

When the Marangoni number is higher than a certain threshold, the basic flow undergoes a transition to a fully 3D, steady-state flow. The numerical results show that the critical Marangoni number $Ma_c$ decreases with increasing Biot number $Bi$, following a power-law (i.e., $Ma_c \sim Bi^{-2/3}$) in the absence of evaporation. Including evaporation leads to a saturation regime wherein $Ma_c$ remains almost unchanged, giving thus a lower bound of the critical Marangoni number ($\approx 120$). Hence, both the conductive heat flux through the gas phase and the evaporative cooling play a destabilizing role, while buoyancy in the liquid layer has a stabilizing effect, though its effect is insignificant. Evaporation-induced destabilizing is due to the amplified thermal gradient in the radial direction (near the sidewall). The numerical results also show that evaporation stabilizes the flow by removing heat from the interface, resulting in a lower interfacial temperature in the central region of the liquid layer. There is a certain amount of competition between stabilizing and destabilizing role played by evaporation. Further research is needed to shed light on this issue. 

\textcolor{black} {The present study showed evidence that the presence of evaporation causes significant cooling of the liquid surface and therefore changes radically the hydrodynamic behavior in the liquid via the Marangoni phenomenon. However, we caution that we have used constant Biot numbers in numerical simulations, which can be a source of high uncertainty when studying these transfer mechanisms, particularly for volatile liquids like FC-72. In that regard, we may follow a numerical study of Gatapova and Kabov~\cite{gatapova2008shear}, in which they proposed a way of calculating the heat transfer coefficient at the liquid-gas interface and obtained numerically the dependence of the Biot number on flow parameters and spatial variables. Similarly, the magnitude of the evaporation rate depends on an uncertain parameter, i.e., the evaporation coefficient, or the evaporative Biot number $B_\mathrm{ev}$, and therefore there is a need of developing reliable methods to estimate this coefficient.}

\section*{Acknowledgements}  This work was financially supported by the National Natural Science Foundation of China (Grants No. 11532015, U1738119),by China's Manned Space Program (TZ-1) and by the Joint Project of CMSA-ESA Cooperation on Utilization in Space, and was conducted within the framework of a joint Doctoral Training Program between Aix-Marseille University and Institute of Mechanics (CAS) with a fellowship provided by the China Scholarship Council (CSC), which is gratefully acknowledged.

 \section*{Conflict of interest}

The authors declare that they have no conflict of interest.

\biboptions{numbers,sort&compress}

\end{document}